\documentclass[showpacs,10pt,aps,prd,reprint,onecolumn,notitlepage,superscriptaddress]{revtex4-1}

\usepackage{amsmath}
\usepackage{amsfonts}
\usepackage{tensor}
\usepackage{graphicx}
\usepackage[]{subfigure}
\usepackage{color}

\usepackage[hidelinks]{hyperref}
\hypersetup{
    bookmarks=true,         
    unicode=true,          
    pdftoolbar=true,        
    pdfmenubar=true,        
    pdffitwindow=true,     
    pdfstartview={FitH},    
    pdftitle={My title},    
    pdfauthor={author},     
    pdfsubject={Subject},   
    pdfcreator={Creator},   
    pdfproducer={Producer}, 
    pdfkeywords={keyword1} {key2} {key3}, 
    pdfnewwindow=true,      
    colorlinks=true ,       
    linkcolor=blue  ,       
    citecolor=blue,      
    urlcolor=black           
}

\providecommand{\ed}{\mathrm{d}}
\providecommand{\m}{\mathrm{m}}
\providecommand{\s}{\mathrm{s}}
\providecommand{\kg}{\mathrm{kg}}
\providecommand{\dim}{\mathrm{dim}}
\providecommand{\E}{\tilde{E}}

\begin{document}

\title{Massive Gravitons on Bohmian Congruences}

\author{Mohsen Fathi}
\email{m.fathi@shargh.tpnu.ac.ir;\,\,\,mohsen.fathi@gmail.com}

\author{Morteza Mohseni}
\email{m-mohseni@pnu.ac.ir}

\affiliation{Department of Physics, 
Payame Noor University (PNU), P.O.\ Box 19395-3697
Tehran, I.R. of IRAN}

\begin{abstract}
Taking a quantum corrected form of Raychaudhuri equation in a geometric background described by a Lorentz-violating massive theory of gravity, we go through investigating a time-like congruence of massive gravitons affected by a Bohmian quantum potential. We find some definite conditions upon which these gravitons are confined to diverging Bohmian trajectories. The respective behaviour of those quantum potentials are also derived and discussed. Additionally, and through a relativistic quantum treatment of a typical wave function, we demonstrate schematic conditions on the associated frequency to the gravitons, in order to satisfy the necessity of divergence.  \\

{\textit{keywords}}: Massive gravity, Quantum corrected Raychaudhuri equation, Time-like congruence expansion

\end{abstract}

\maketitle

\section{Introduction}
The  natural inclinations of alternative theories of gravity, beside capturing those solar system (and standard cosmic) predictions of general relativity, is to generate a self contained theory of gravitation, applicable to comprehend the universe without relying on dark energy or dark matter. However talking about matter, somehow arises the necessity of matter-gravity coupling which has not been yet achieved in the shape of a fully re-normalizable theory of gravitation. To do so, such theory may become impregnated by massive gravitational quanta (gravitons), not unlike that of Fierz and Pauli in 1939 \cite{Fierz1939}. The Lorentz invariant Fierz-Pauli massive spin-2 theory however appears to suffer some shortcomings, preventing it of becoming a reliable massive gravity theory; like strong coupling in infrared ranges \cite{Arkani-Hamed2003} which necessitates a UV completion to be presumable within small ranges like the Solar system.  In fact it is well-known that if gravitation were to describe massive gravitons of mass $m$, then the resultant perturbations would be so small and negligible inside the Compton wavelength $\lambda_c\sim\frac{\hbar}{mc}$, with $c$ to be the speed of light. Intuitively, the mass of gravitons are included through adding a term to the gravitational action, implying non-zero mass in the linearized level. However what is mostly observed in massive theories of gravity is that the gravitons become massive through the Higgs mechanism, under the impact of scalar fields induced in a gauge-invariant gravitational action \cite{Hooft2007, Dubovsky2004,Rubakov2008}. However besides the above Lorentz-invariant theories, the Lorentz-violating massive gravity theories have also appeared to be in the focus of a large amount of research and discussions. In these theories, the mass is acquired through the spontaneous symmetry breaking (for very expressive notes on Lorentz-invariant/violating massive theories please see Ref.s~\cite{Bebronne2009, de Rham2014}). \\

In this paper we also consider a Lorentz-violating massive theory, and now that these theories do assign non-zero mass to the gravitational particles, it makes sense to investigate the way that such particles move on time-like trajectories on the geometric background described by the theory. Indeed one can consider the time-like gravitons moving on black holes, formed in massive gravity. Such black holes were discussed in Ref.s~\cite{Comelli2011, Fernando2014} and also the effective potentials and their corresponding orbits of test massive particles moving on a Lorentz-violating massive gravity black hole has investigated in Ref.~\cite{Zhang2015}. However the very concept of a typical black hole is usually granted by the formation of a singularity, mostly discussed on the premise of Hawking-Penrose singularity theorems \cite{Penrose1965, Hawking1970, Hawking1973}. Traditionally, such theorems are approached by the time-like and null-like combinations of Raychaudhuri's kinematical decomposition and his famous equation \citep{Raychaudhuri1955}. Indeed the Raychaudhuri equation is also of great significance in treating the (an)isotropic cosmological models based on a vorticity-free fluid, and also performs a key role in examining the focusing of a congruence of particle trajectories while falling onto singular regions such as black holes. In the case of massive gravity however, those time-like congruences falling onto black holes are essentially consisting of massive gravitons. In this regard, the congruence basically holds a mixture of quantum particles moving in a gravitational field. A way to dealing with this, goes through the Bohmian behaviour of quantum particles in a curved spacetime, while they are affected by a Bohmian quantum potential (for a very good review see Ref.~\cite{Licata2014}). Accordingly, a congruence of massive gravitons can be decomposed into spacetime and quantum kinematical characteristics. To include this, one has to modify the Raychaudhuri equation in order to contain both of these categories. It was Das \cite{Das2014}, who applied such a modification and later it was shown that this modified version results in the correctly  evaluated graviton mass in the limit of the associated Compton wave length \cite{Ali2015}. Therefore it makes sense to regard time-like trajectories in massive theories of gravity, as those under the influence of a Bohmian quantum potential, and decompose them by means of a quantum corrected Raychuadhuri equation (QRE). In this paper we also consider such a situation in the context of a Lorentz-violating massive theory of gravity. These steps are tackled: In Sec.~\ref{sec:time-like general} we calculate a radially propagated time-like congruence of massive particles in a general spherically symmetric background spacetime geometry and impose the results in the QRE to obtain a general framework for the quantum potential. In Sec.~\ref{sec:Massive Gravity} we consider the special case of a Lorentz-violating massive theory of gravity and exploit its spherically symmetric solution in order to specify the quantum potential for two different cases of study. There, and for each case, we discuss the energy conditions upon which the graviton congruence can avoid any geodesic incompleteness (focusing). Such condition is crucial to Bohmian trajectories. Determination of such quantum potentials helps us to study each case separately. Furthermore and in Sec.~\ref{sec:frequency}, we switch to a Klein-Gordon type equation to rule a non-geodesic congruence of gravitons, bounded by a quantum potential. This way we can replicate the procedure of finding divergence, in a way that instead of geodesic equations, the trajectories are essentially emerge from their confinement to the Bohmian quantum potential. We apply this to find conditions on the associated frequency to the gravitons. We summarize in Sec.~\ref{sec:conclusion}. Throughout this paper we denote overdots for differentiation with respect to the congruence parametrization and primes for differentiation with respect to radial coordinate. Also 4-dimensional indices are presented by $a, b, c, ...$, whereas those 3-dimensional ones, by $i, j$.

\section{Radially Propagated Time-Like Congruence in a Quantum Potential}\label{sec:time-like general}

Assume a congruence of massive particles falling onto a region, geometrically described by the line element
\begin{equation}\label{eq:metric-general}
-c^2\ed\tau^2=-f(r)c^2\ed t^2+\frac{1}{f(r)}\ed r^2+r^2\ed\Omega_{(2)}^2,
\end{equation}
with $\tau$ to be the proper time and here, the congruence parametrization. Now if a radially propagated time-like congruence is generated by a time-like vector field $u^a=(u^0,u^1,0,0)$, then it is governed by the condition $-c^2=g_{ab}u^a u^b$, characterizing the light cone structure by
\begin{equation}\label{eq:timelike condtion-general}
-c^2=-f(r)(u^0)^2+\frac{1}{f(r)}(u^1)^2.
\end{equation}
Defining the 4-momentum $p^a=m u^a$, then it is straightforward to write down the Hamilton-Jacobi equation
\begin{equation}\label{eq:H-J-general}
g_{ab}p^ap^b+m^2c^2=0.
\end{equation}
The motion of massive particles on such congruence can be also expressed in terms of the particles' energy $E$, as a constant of motion. Indeed one can define $E\doteq-p_0 c$ \cite{Misner1973}, giving $u^0=\frac{\E c}{f(r)}$ with $\E=E/mc^2$ to be a dimensionless ratio, illustrating that to what extent the energy of a test particle can approach to that in the asymptote. Accordingly, Eq.~(\ref{eq:timelike condtion-general}) yields
\begin{equation}\label{eq:u-general}
u^a=\left(\frac{\E c}{f(r)}, -c\sqrt{\E^2-f(r)},0 ,0\right).
\end{equation} 
It is straightforward to show that $u^a$ is parallel-transported along the congruence it generates ($u\indices{^a_{;b}}u^b=\mathbf{0}$), so the congruence is a geodesic congruence. Now if we consider a cross-sectional 3-surface, transverse to the congruence generated by $u^a$, then one can define the transverse expansion \cite{Poisson2004}
\begin{equation}\label{eq:expansion-general}
\Theta=h^{ab}u_{a;b},
\end{equation}
in which 
 \begin{equation}\label{eq:h-general}
h^{ab}=g^{ab}+\frac{u^a u^b}{-u^c u_c}
\end{equation}
is the metric describing the cross-sectional surface. Accordingly, for a radially propagated Bohmian congruence of massive gravitons (axions) under the influence of the quantum potential $V_Q$, the transverse expansion evolution is governed by the rotation-less QRE, which in the $(-,+,+,+)$ sign convention is of the form \cite{Das2014}
 \begin{equation}\label{eq:QRE-general}
\dot\Theta+\frac{1}{3}\Theta^2=-\sigma^2-R_{ab}u^au^b+h^{ab}\partial_b\partial_a V_Q+\frac{\epsilon_1\hbar^2}{m^2}h^{ab}\partial_b\partial_a R.
\end{equation}
In the above equation, $\sigma^2=\sigma_{ab}\sigma^{ab}$ where $\sigma_{ab}=u_{(a;b)}-\frac{1}{3}h_{ab}\Theta$ is the symmetric trace-less shear tensor. Also $R_{ab}$ is the Ricci tensor and $R=R\indices{^a_a}$. The congruence avoids geodesic incompleteness (convergence or focusing) if $\dot\Theta+\frac{1}{3}\Theta^2>0$ \cite{Poisson2004}. Moreover the dimensionless coefficient $\epsilon_1$ is supposed to be $1/6$ for conformally invariant theories describing massless particles, whereas for massive Dirac particle it is $1/4$ \cite{Das2014}. The quantum potential is indeed the rate of change of the amplitude $\mathcal{R}$ of an associated flat wave function $\psi=\mathcal{R}\textrm{e}^{\textrm{i}S}$ with respect to the background spacetime and has been essentially defined by $V_Q=\frac{\hbar^2}{m^2}\left(\frac{\square\mathcal{R}}{\mathcal{R}}\right)$. However Eq.~(\ref{eq:QRE-general}) itself provides a peculiar expression for the quantum potential, emerged directly from the background geometry. Also regarding the radial nature of the congruence, it is intuitive to assume $V_Q\equiv V_Q(r)$. Therefore recalling $\dot\Theta=\Theta_{,a}u^a $, and using Eq.s~(\ref{eq:metric-general}), (\ref{eq:u-general}), (\ref{eq:expansion-general}) and (\ref{eq:h-general}), we can infer 
an appropriate equation governing the behaviour of the quantum potential on a spherically symmetric background in Eq.~(\ref{eq:metric-general}). We will do this in the next section to derive specific quantum potentials in the case that the metric potential is expressed in a massive theory of gravity. We continue our discussion by bringing essentials of a Lorentz-violating massive theory and its static spherically symmetric solution, and investigate the radially propagated massive gravitons in two distinct cases.

\section{Massive Gravitons on Bohmian Trajectories }\label{sec:Massive Gravity}

The Lorentz-violating massive gravity action we consider in this paper is \cite{Dubovsky2004, Rubakov2008, Bebronne2009}
 \begin{equation}\label{eq:action}
S=\int\ed^4x~\sqrt{|g|}\left[\frac{R}{16\pi}+\Lambda^4\mathcal{F}(X,W^{ij})\right],
\end{equation}
in which $X$ and $W^{ij}$ are functions of the Goldstone fields $\Phi^0$ and $\Phi^i$ and are written as
\begin{subequations}\label{eq:X,W-Goldstone} 
\begin{align}
X&=\frac{g^{ab} \partial_a\Phi^0\partial_b\Phi^0}{\Lambda^4},\label{eq:X,W-Goldstone-X}
\\
W^{ij}&=\frac{g^{ab} \partial_a\Phi^i\partial_b\Phi^j}{\Lambda^4}-\frac{g^{cd}g^{ef} \partial_c\Phi^i\partial_d\Phi^0 \partial_e\Phi^j\partial_f\Phi^0}{\Lambda^4X}.\label{eq:X,W-Goldstone-W}
\end{align}
\end{subequations}
Here, $\Lambda=\left(m ~\mathcal{M}_{\textrm{\tiny{Planck}}}\right)^{\frac{1}{2}}$, with $\mathcal{M}_{\textrm{\tiny{Planck}}}$ as the Planck mass.  Presented in Ref.s~\cite{Bebronne2009b, Comelli2011}, the corresponding vacuum spherically symmetric solution to the resultant field equations is of the form
\begin{equation}\label{eq:solution}
f(r)=1-M\left(\frac{2}{r}+\frac{\gamma}{r^\lambda}\right),
\end{equation}
in which, since $f(r)$ in Eq.~(\ref{eq:metric-general}) has to be dimensionless, therefore in SI units $\dim[M]=\m$ and $\dim[\gamma]=\m^{\lambda-1}$ ($\m$ represents meters). Note that defining $Q\doteq\gamma M$, we get back to what which was used in Ref.s~\cite{Fernando2014, Zhang2015}, however the adoption of the form in Eq.~(\ref{eq:solution}) becomes more beneficial within this text. Accordingly, Eq.~(\ref{eq:QRE-general}) yields 
\begin{equation}\label{eq:VQ-equation-general}
  \begin{aligned}
  r^{\lambda +3} \left[2 \E^2 r^{\lambda +2} V_Q''+ \left(M \left(\gamma  (\lambda -4) r-6 r^{\lambda }\right)+4 r^{\lambda +1}\right)V_Q'\right]\\
  ={\gamma  \left(\lambda ^3-\lambda ^2-4 \lambda +4\right) M {\epsilon_1} \frac{\hbar^2}{m^2} \left[M \left(\gamma  (\lambda -4) r-6 r^{\lambda }\right)-2 \left(\E^2 (\lambda +3)-2\right) r^{\lambda +1}\right]}.
  \end{aligned}
\end{equation}
It is worth noting that $\dim[V_Q]=\m^2 \s^{-2}$ \cite{Das2014}. In what follows, we consider two special cases by specifying $\lambda$ in the above equation and discuss the conditions of congruence divergence. Note that to avoid generating an infinite Arnowitt-Deser-Misner (ADM) mass, one should always adopt $\lambda\geq1$, which also guarantees flatness of $f(r)$ in the asymptote. We investigate the cases of $\lambda=1$ and $\lambda=4$.

\subsection{The Case of $\lambda=1$}\label{subsec:lambda=1}

\begin{figure}
\center{ 
\includegraphics[width=7cm]{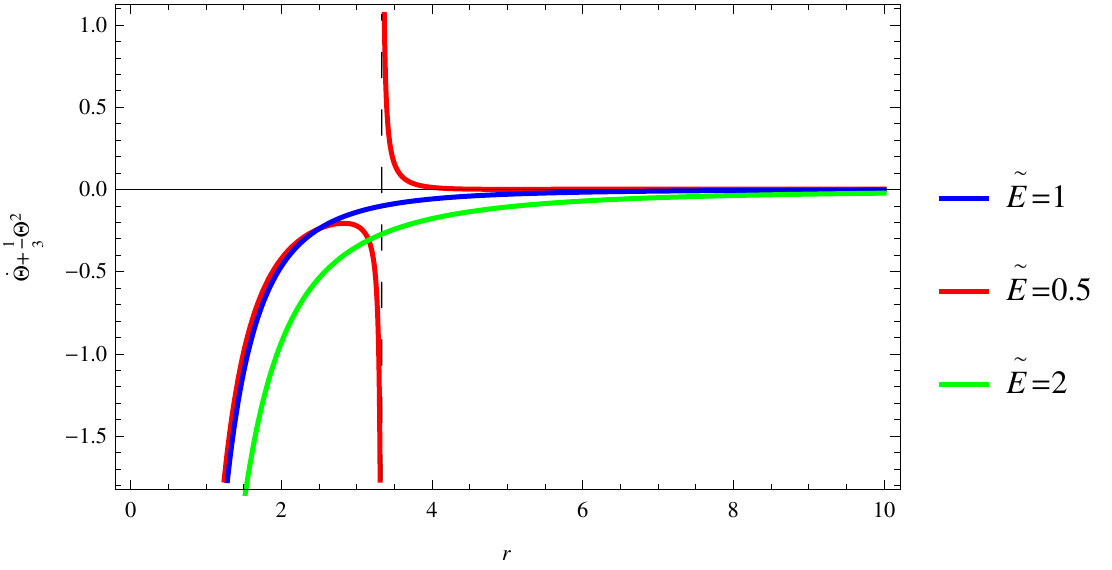}~(a)
\hfil
\includegraphics[width=7cm]{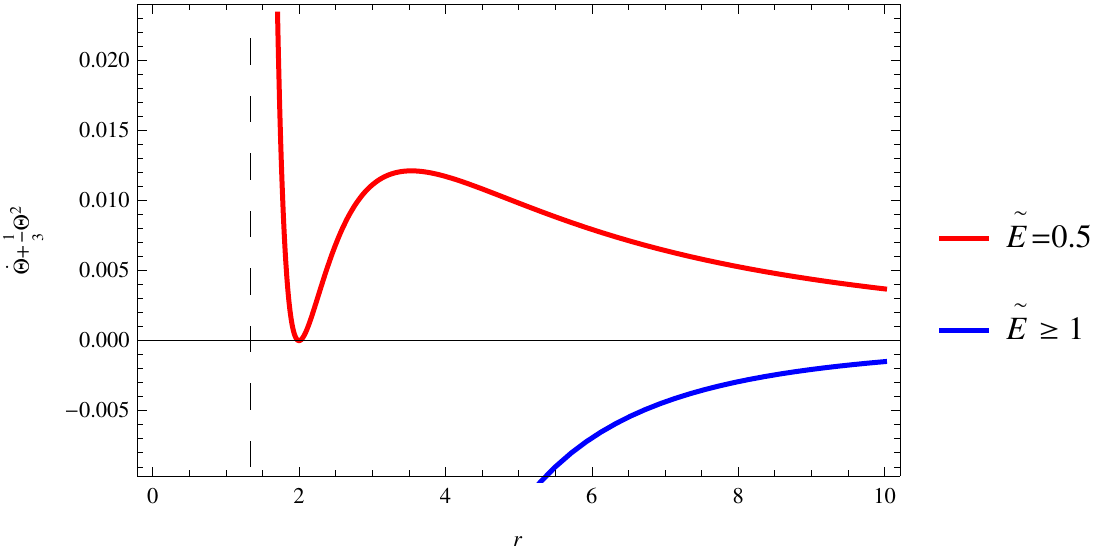}~(b)
\hfil
\includegraphics[width=7cm]{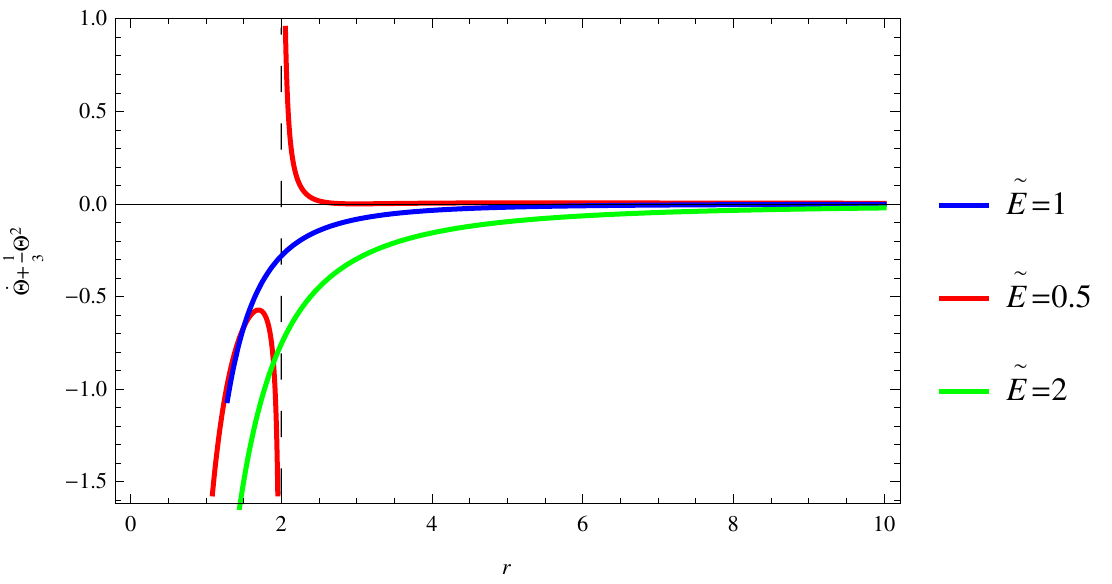}~(c)
\caption{\label{fig:divergence-lambda=1} The behaviour of $\dot\Theta+1/3\Theta^2$ for $\lambda=1$ and different $\E$. The  evaluations have been done for (a) $\gamma>0$, (b) $\gamma=-1$ and (c) $-1<\gamma<0$. We can see that only in the case of $\E=1/2$ and $r>r_1$ we may encounter positivity. }}
\end{figure}

\begin{figure}
\center{  \includegraphics[width=8cm]{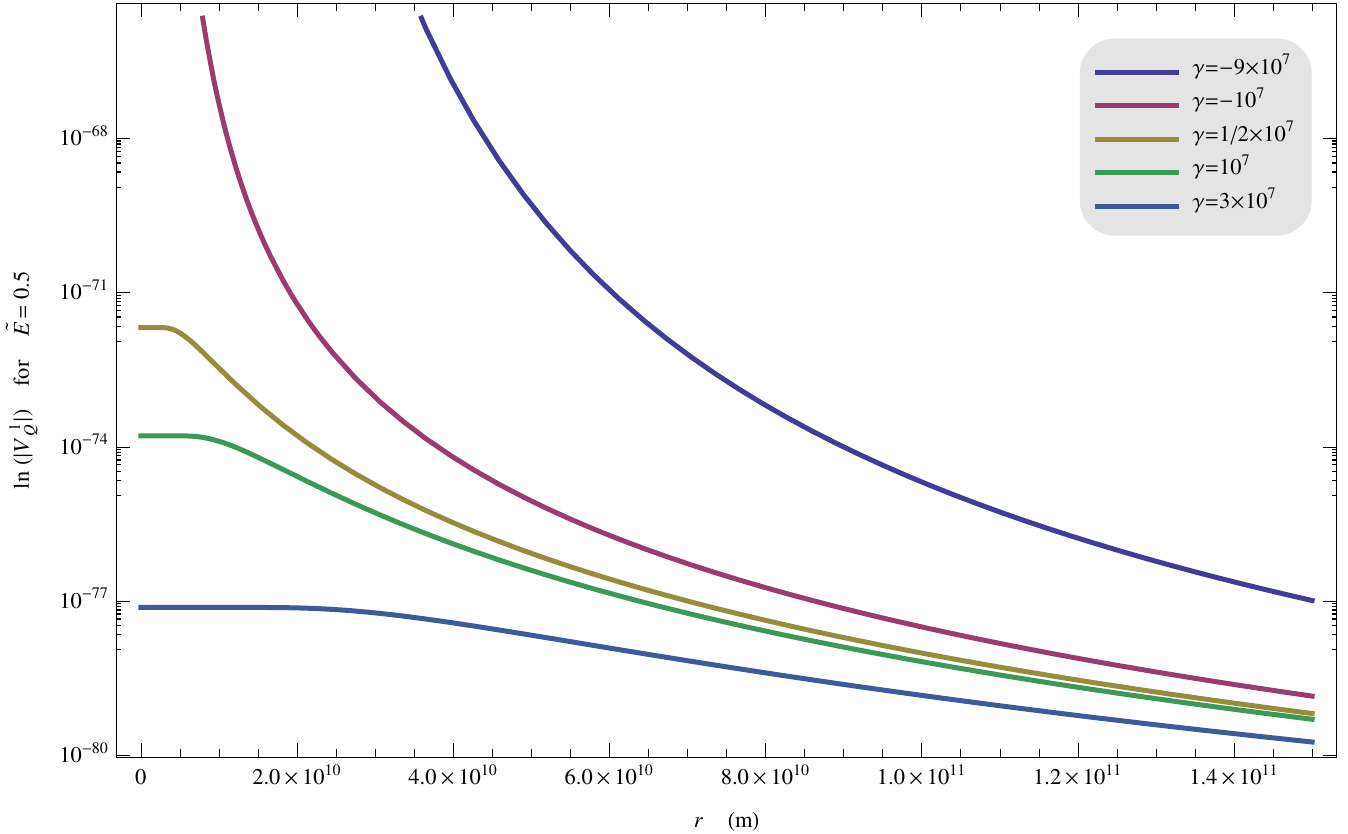}
\caption{\label{fig:VQ-lambda=1} The logarithmic behaviour of $|V_Q^1|$ for different values of $\gamma$, according to a congruence of gravitons falling onto a massive object like the Sun. The radial distance has been put in a way that every measurements are done within the separation between the Sun and the Earth. Since the results are rather small, we had to consider large values of $\gamma$ in order to obtain visible distinctions among the curves.}}
\end{figure}

In this case, solving Eq.~(\ref{eq:VQ-equation-general}) we get
\begin{equation}\label{eq:VQ-lambda=1}
  \begin{aligned}
 V^1_Q(r)= c_2-\frac{c_1(\gamma +2)}{\E^2}  \left(\frac{3}{2}\right)^{1-\frac{2}{\E^2}} M r^{-\frac{2}{\E^2}}\\ \times\left[r^{\frac{2}{\E^2}} \left(\frac{\E^2}{\gamma  M+2 M}\right)^{\frac{2}{\E^2}} \Gamma \left(\frac{2}{\E^2}-1,\frac{3 M (\gamma +2)}{2 \E^2}\right)-\left(\frac{\E^2 r}{\gamma  M+2 M}\right)^{\frac{2}{\E^2}} \Gamma \left(\frac{2}{\E^2}-1,\frac{3 M (\gamma +2)}{2 \E^2 r}\right)\right],
  \end{aligned}
\end{equation}
in terms if incomplete Gamma functions. Here $c_1$ and $c_2$ are integration constants. However as we have stated before, the divergence condition ($\dot\Theta+1/3\Theta^2>0$ which is naturally equivalent to the positiveness of the r.h.s. of Eq.~(\ref{eq:QRE-general}) for consistent expressions for $V_Q$) must be applied to find conditions on $\E$, in order to obtain permanent divergence of the Bohmian congruence of gravitons. Therefore it is of benefit to obtain a visual sense about the behaviour of $\dot\Theta+1/3\Theta^2$ in the spacetime described by Eq.~(\ref{eq:solution}). For the case of $\lambda=1$, this scalar is 
\begin{equation}\label{eq:thetadot-lambda=1}
 \left(\dot\Theta+\frac{1}{3}\Theta^2\right)|_{\lambda=1}=-\frac{c^2 \left[2 \left(\E^2-1\right) r+3 (\gamma +2) M\right]^2}{6 r^3 \left[\left(\E^2-1\right) r+(\gamma +2) M\right]}.
\end{equation}
This behaviour for different values of $\gamma$ and $\E$ has been shown in Fig.~\ref{fig:divergence-lambda=1}. The main aim is the determination of some edges for $\E$. However since $\E$ is dimensionless, to plot the expression in Eq.~(\ref{eq:thetadot-lambda=1}) we tentatively adopt $c=1$. This means that we have discarded dimensionality for this very peculiar moment, to provide desired values of the ratio $\E$ within a small radial region, which is here $r\in[0,10]$. We have also let $M=1$, without loose of generality. It is interesting knowing that in all cases (even for other values of $M$), we find out that the only reasonable range of energy to have permanent divergence, would be $0<\E<1$. Moreover, as we can observe in Fig.~\ref{fig:divergence-lambda=1}(b), there may be points at which $\dot\Theta+1/3\Theta^2\leq0$, implying congruence convergence. In this regard and in order to avoid geodesic incompleteness, we should consider the regions where $r>r_1$ with 
\begin{equation}\label{eq:r1}
r_1=\frac{3}{2}\frac{M(2+\gamma)}{1-\E^2},
\end{equation}
which lies within physical radial distances for $\gamma\geq-2$. This also includes regions beyond the asymptote.
Therefore, a reasonable choice for $V_Q^1(r)$ is obtained by inserting $\E=1/2$ in Eq.~(\ref{eq:VQ-lambda=1}), giving
\begin{equation}\label{eq:VQ-lambda=1-half}
V_Q^1(r)|_{\E=1/2}=c_2-\frac{c_1 \left[\Gamma (7,6 M (\gamma +2))-\Gamma \left(7,\frac{6 M (\gamma +2)}{r}\right)\right]}{279936 (\gamma +2)^7 M^7}.
\end{equation}
Surely $\dim[c_1]=\frac{\m^9}{\s^2}$ and $\dim[c_2]=\frac{\m^2}{\s^2}$. One can also inspect the behaviour of the quantum potential while the graviton congruence recedes from the spherical massive source $M$. To deal with this, we should note that since $\mathcal{R}$ is only valid inside the associated Compton wavelength of gravitons ($\lambda_c\sim10^{26}\m$, which is equal to the radius of the observable universe) therefore it is plausible to think of a vanishing potential at infinity, resulting in
\begin{equation}\label{eq:VQ-lambda=1-half-without c2}
V_Q^1(r)|_{\E=1/2}=\frac{c_1 \left[\Gamma \left(7,\frac{6 M (\gamma +2)}{r}\right)-720\right]}{279936 (\gamma +2)^7 M^7}.
\end{equation}
This means that $V^1_Q$ is just being rescaled by a simple coefficient $c_1$. Therefore to compare the behaviours for different values of $\gamma$ we can just rule out $c_1$ (equivalently letting $c_1=1$). Fig.~\ref{fig:VQ-lambda=1} shows how the quantum potential changes while we recede from a spherical massive object (here considered to be the Sun of mass $M\approx 2\times10^{30}\kg=1482.22\m$ with the distance $r\approx 150\times10^9\m$ from the Earth). According to the figure, $V_Q^1$ is available around the massive object and it decreases from its maximum values at regions near the massive object, to its minimum values while the distance is increased (here we considered regions near the Earth). Note that the values of $V_Q^1$ are very small, even by considering rather large values of $\gamma$ (either positive or negative).

\subsection{The Case of $\lambda=4$}\label{subsec:lambda=4}

\begin{figure}
\center{  \includegraphics[width=7cm]{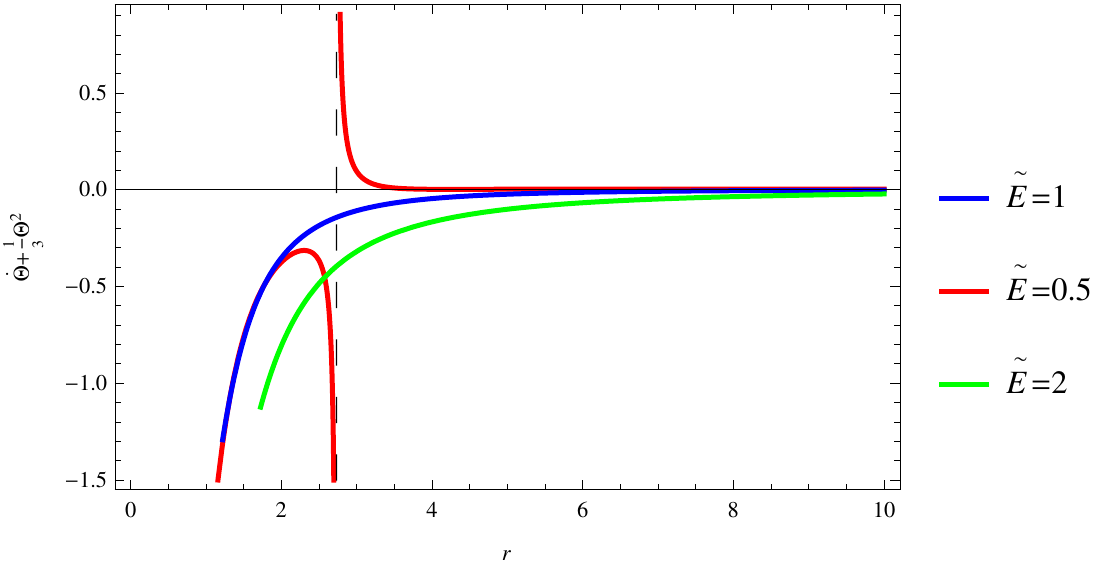}~(a) \hfil
\includegraphics[width=7cm]{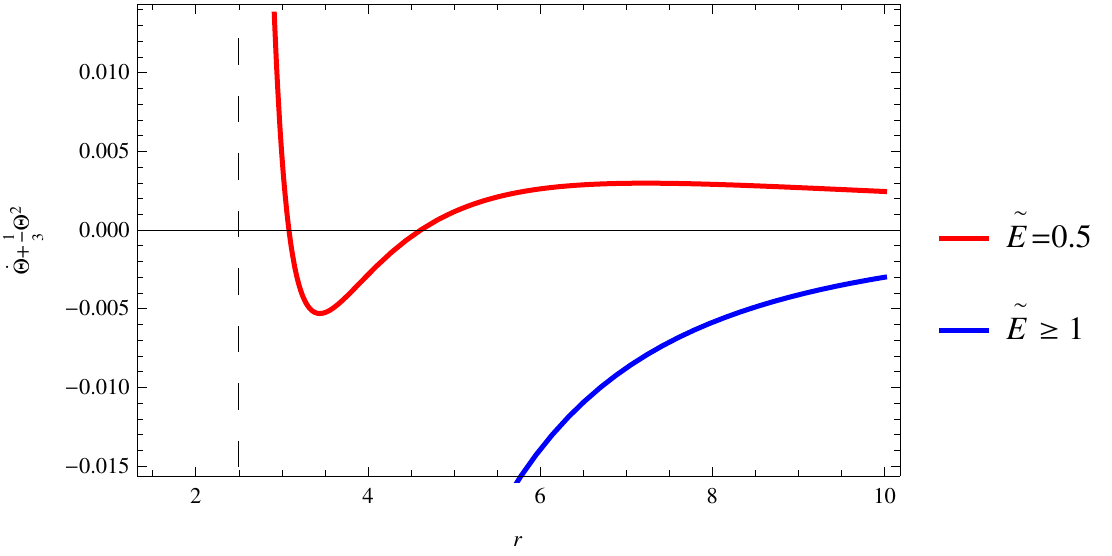}~(b)
\caption{\label{fig:divergence-lambda=4} The behaviour of $\dot\Theta+1/3\Theta^2$ for $\lambda=4$ and different $\E$. The  evaluations have been done for (a) $\gamma>0$ and (b) $\gamma<0$. We can see that only in the case of $\E=1/2$ and $r>r_4$ we may encounter positivity. }}
\end{figure}

\begin{figure}
\center{  \includegraphics[width=8cm]{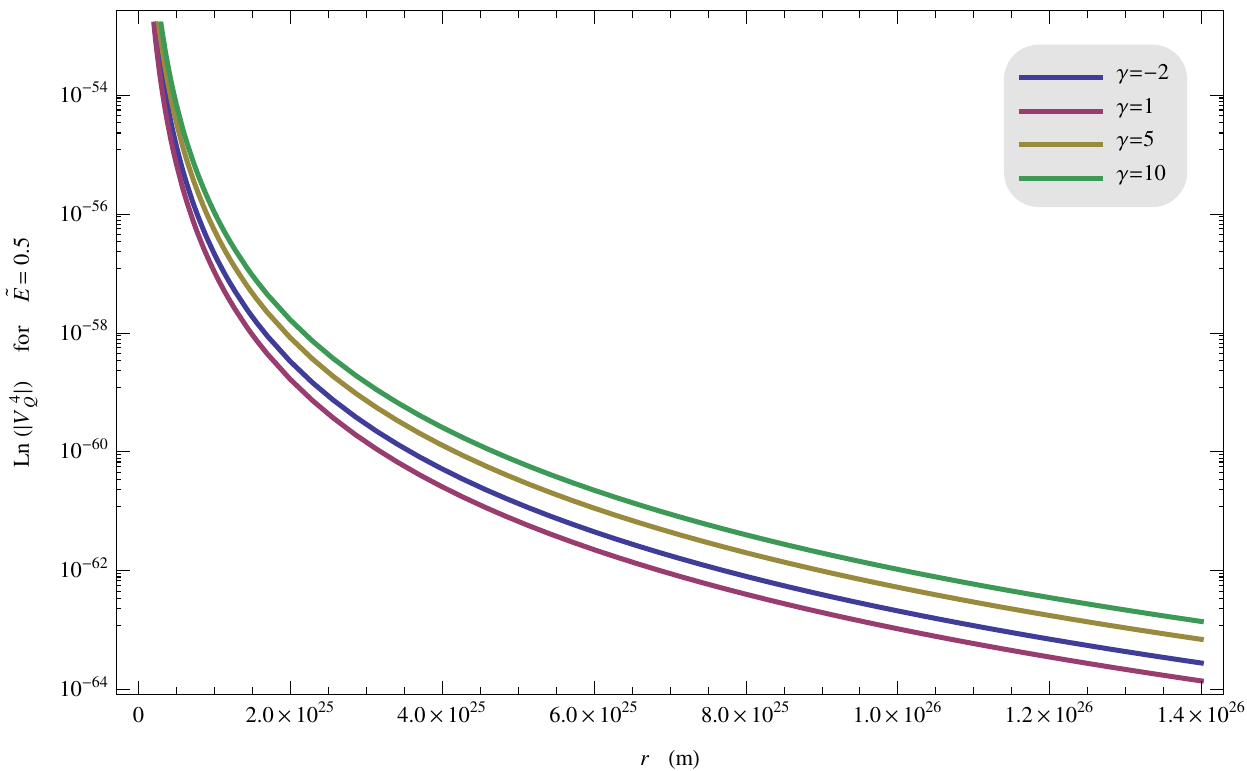}
\caption{\label{fig:VQ-lambda=4} The logarithmic behaviour of $|V_Q^4|$ for different values of $\gamma$, according to a congruence of gravitons moving on the edge of the observable universe. In this regard it is affected by the total mass, within the universe. }}
\end{figure}

In this case Eq.~(\ref{eq:VQ-equation-general}) provides
\begin{equation}\label{eq:VQ-lambda=4}
V_Q^4(r)=c_2-\frac{6 \gamma  M \epsilon_1 \hbar ^2}{m^2 r^6}+c_1   \left(\frac{3M}{\E^2 }\right)^{1-\frac{2}{\E^2}} \Gamma \left(\frac{2}{\E^2}-1,\frac{3 M}{\E^2 r}\right).
\end{equation}
As in the previous case, to satisfy the divergence condition we should obtain some appropriate intervals for $\E$. Dealing with the expression
\begin{equation}\label{eq:thetadot-lambda=4}
 \left(\dot\Theta+\frac{1}{3}\Theta^2\right)|_{\lambda=4}=-\frac{c^2 \left[4 r^8 \left[\left(\E^2-1\right) r+3 M\right]^2-12 \gamma  \left(\E^2-1\right) M r^6\right]}{6 r^7 \left[\left(\E^2-1\right) r^5+M r \left(\gamma +2 r^3\right)\right]},
\end{equation}
we can adopt all previous conditions to seek for permanent divergence. According to Fig.~\ref{fig:divergence-lambda=4} it turns out that it is still $0<\E<1$ which can guarantee this. 
Moreover, to avoid convergence in the case of $\gamma<0$, we must also consider $r>r_4$ in which
\begin{equation}\label{eq:r4}
r_4=\frac{\left[{4 \sqrt{3M} \left(\E^2-1\right) \sqrt{\gamma \left( \E^2 -1  \right)}+9 M^2}\right]^\frac{1}{2}-3 M}{2 \left(\E^2-1\right)}
\end{equation} 
eliminates $\dot\Theta+1/3\Theta^2$. Therefore a valid quantum potential, consistent with a diverging Bohmian congruence in the case of $\lambda=4$ could be
\begin{equation}\label{eq:VQ-lambda=4-half}
V_Q^4(r)|_{\E=1/2}=\frac{c_1 \Gamma \left(7,\frac{12 M}{r}\right)}{35831808 M^7}+c_2-\frac{6 \gamma  M \epsilon_1 \hbar ^2}{m^2 r^6},
\end{equation}
which is indeed of a huge value, since for gravitons $\frac{\hbar}{m}\sim10^{34}\frac{\m^2}{\s}$ \cite{Goldhaber2010}. Once again since we consider $V_Q^4$ to vanish at infinity, we can recast the above expression in the form
\begin{equation}\label{eq:VQ-lambda=4-half-without c2}
V_Q^4(r)|_{\E=1/2}=\frac{c_1 \left(\Gamma \left(7,\frac{12 M}{r}\right)-720\right)}{35831808 M^7}-\frac{6 \gamma  \eta ^2 M \epsilon_1}{r^6},
\end{equation}
with $\eta=\frac{\hbar}{m}$. Since $V_Q^4$ is extremely large within spacial separations like the Solar system, it makes sense to examine it in a much broader sense. The diagram in Fig.~\ref{fig:VQ-lambda=4} shows the behaviour of this quantum potential when the graviton congruence is moving at the threshold of the observable universe of mass $M\approx7\times10^{24}\m$ \cite{Davies2008} and radial separation $r\approx1.4\times10^{26}\m$ \cite{Bars2009}.

\section{Relativistic Quantum Treatment of Divergence}\label{sec:frequency}

Once the particles' confinement to quantum potentials has to be taken into account at a very initial step, it is inevitable to replace the geodesic trajectories by those which are obtained from a bounded Hamilton-Jacobi equation. This is indeed the approach which has been considered as a conceptual foundation to derive the QRE.  In this section, we go through this approach to reconcile our treatment onto the behaviour of non-geodesic graviton congruence. Therefore, and to obtain appropriate conformity with the usual formulation of relativistic quantum mechanics in the literature and also with Ref.~\cite{Das2014}, we choose the $(+, -,-,-
)$ sign convention throughout this section (the metric associated to Eq.~(\ref{eq:metric-general}) becomes $g_{ab}=\textrm{diag}\left(f(r),-f(r)^{-1},-r^2,-r^2\sin^2\theta\right)$). So, for a time-like tangential vector $v^a$ it is mandatory to satisfy  $v^a v_a>0$.\\

\begin{figure}
\center{  \includegraphics[width=8cm]{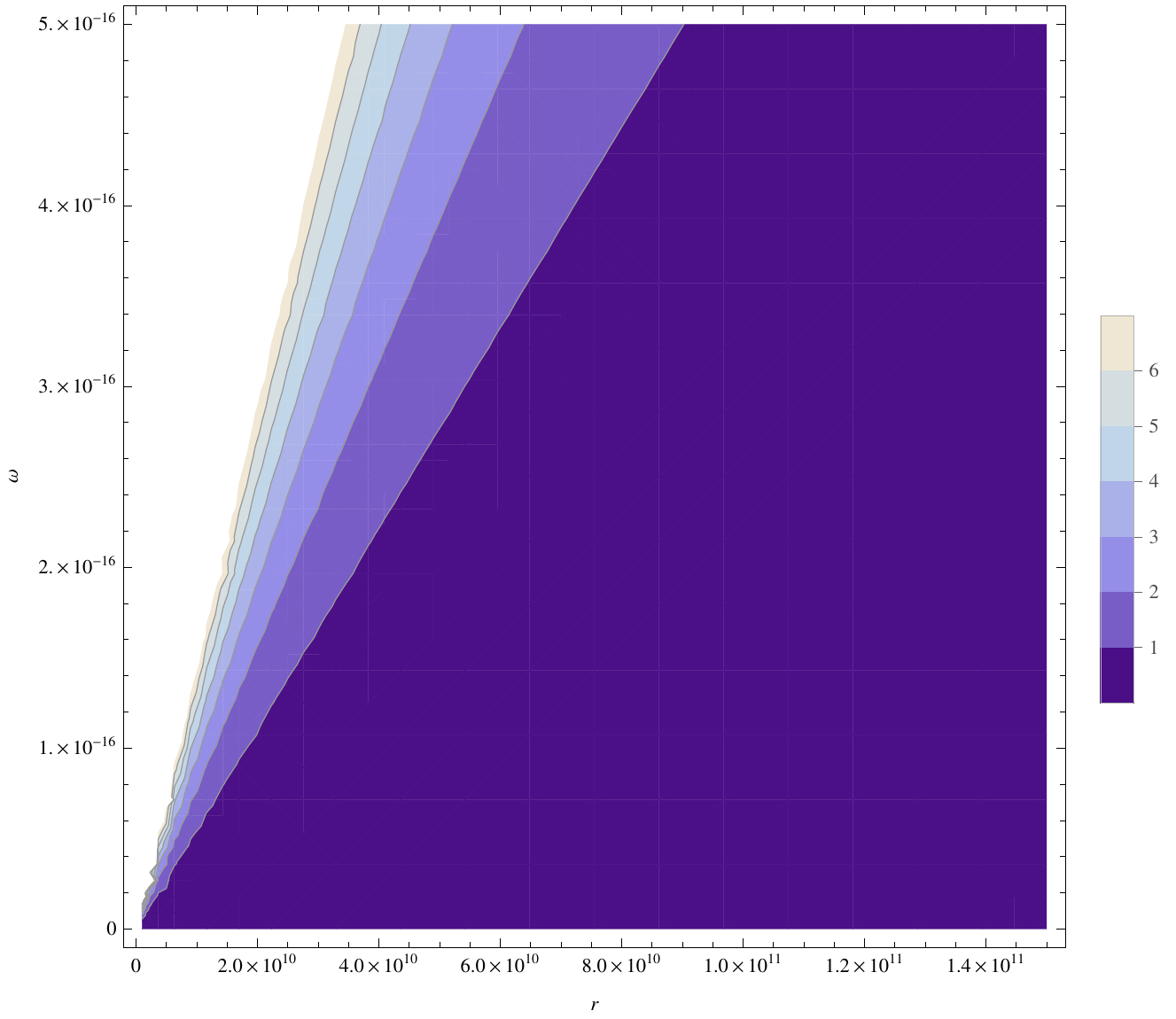}
\caption{\label{fig:Quantum-lambda=1} The behaviour of $\dot\Theta+\frac{1}{3}\Theta^2$ for $\lambda=1$, $\gamma=3$ and for very small values of $\omega$ in the solar system constraint. One can see that even in these small ranges of energy, the massive particles will always reside on diverging congruences.}} 
\end{figure}

\begin{figure}
\center{  \includegraphics[width=8cm]{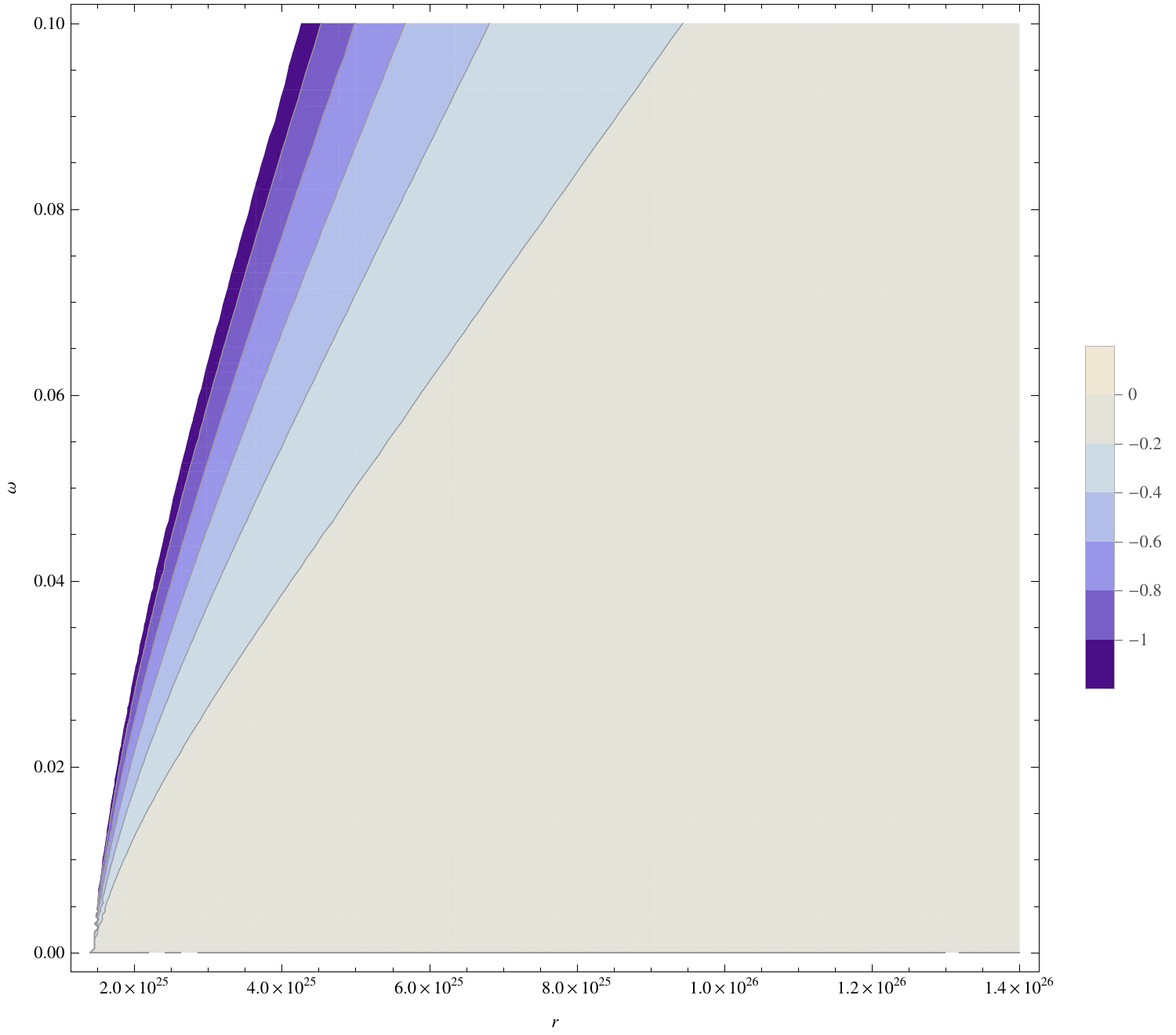}
\caption{\label{fig:Quantum-lambda=4} The behaviour of $\dot\Theta+\frac{1}{3}\Theta^2$ for $\lambda=4$, $\gamma=3$ at the particle horizon threshold. The particles need much greater values of energy than that in the $\lambda=1$, however convergence is inevitable}} 
\end{figure}

First of all, let us define a radial wave vector 
\begin{equation}\label{eq:waveVector-general1}
k_a=\left(k_0,k_r,0,0\right)
\end{equation}
to describe bosons. If $k_a$ is supposed to relate to the energy-momentum of the particle, then it is reliable to consider
\begin{equation}\label{eq:waveVector-general2}
k_a=\left(\frac{\omega}{c},k_r,0,0\right),
\end{equation}
with $\omega$ as the associated frequency to bosons. Note that $p_0=\hbar k_0=\frac{\hbar\omega}{c}$, and therefore $E=p_0c=\hbar\omega$ is the energy of the particles. However to determine $k_r$, one should apply the Hamilton-Jacobi condition, which in relativistic quantum mechanics provides the Klein-Gordon equation. Here on the other hand, and since the bosons are supposed to bounded, this equation is recast in the form \cite{Das2014}
\begin{equation}\label{eq:Klein-Gordon}
\left(\Box+\frac{m^2 c^2}{\hbar^2}-\epsilon_1 R\right)\psi(x^a)=0,
\end{equation}
which same as before, $\psi(x^a)$ is the wave function on a fixed background. Such wave function describes a quantum fluid or a condensate. Substitution of its definition in Eq.~(\ref{eq:Klein-Gordon}) gives
\begin{equation}\label{eq:Condition-VQ}
g^{ab}k_a k_b-\frac{m^2 c^2}{\hbar^2}+\epsilon_1 R-\frac{m^2}{\hbar^2}V_Q=0,
\end{equation}
to obtain which, we should assume the operator form of the wave vector; i.e. $\hat k_a=\mathrm{i}\partial_a$. In this regard, the Klein-Gordon (or the Hamilton-Jacobi) equation would be $\left(\hat k^a\hat k_a\right)\psi=\left(\frac{m^2 c^2}{\hbar^2}-\epsilon_1 R\right)\psi$. If solved for $V_Q$, the above equation yields
\begin{equation}\label{eq:VQ-new}
V_Q(r)=\frac{1}{c^2 m^2 r^2 f(r)}\left[c^2 f(r) \left(-c^2 m^2 r^2+r^2 {\epsilon_1 } \hbar ^2 f''(r)+4 r {\epsilon_1} \hbar ^2 f'(r)-2 {\epsilon_1} \hbar ^2\right)+c^2 \hbar ^2 f(r)^2 \left(2 {\epsilon}_1 k_r^2 r^2\right)+r^2 \omega ^2 \hbar ^2\right].
\end{equation}
Moreover, the relevant non-geodesic congruence is generated by the tangential 4-vector $v_a=\frac{\hbar k_a}{m}$, obeying 
\begin{equation}\label{eq:geodesic-modified}
v^b{}_{;a}v^a=-g^{bc}\left[\frac{\epsilon_1\hbar^2}{m^2}\partial_c\left(R-V_Q\right)\right].
\end{equation}
Calculations in the new sign convention and considering Eq.s~(\ref{eq:waveVector-general2}) and (\ref{eq:VQ-new}), the above condition results in 
\begin{equation}\label{eq:kr}
k_r=\pm\left(\frac{2 c^2 {\epsilon_1} f^2 \left(m^2-{\epsilon_1} \hbar ^2\right) \left(r^3 f'''+4 r^2 f''-2 r f'+4\right)+8 c^2 {\epsilon_1} f^3 \left({\epsilon_1} \hbar ^2-m^2\right)-\omega^2r^3  f' \left(m^2-2 {\epsilon_1} \hbar ^2\right)}{c^2 r^3 f^2 f'\left(m^2-2 {\epsilon_1} \hbar ^2\right)}\right)^{\frac{1}{2}},
\end{equation}
which we consider the negative segment to recover the ingoing congruence. Now to deal with the indigenous constant $\dot\Theta+\frac{1}{3}\Theta^2$, we should note that the projection tensor is essentially based on a 3-metric, which is
\begin{equation}\label{eq:3-metric}
h_{ab}=g_{ab}-\frac{v_a v_b}{v^c v_c},
\end{equation}
where $v_a$ can be extrapolated from Eq.s~(\ref{eq:waveVector-general2}) and (\ref{eq:kr}). Using Eq.s~(\ref{eq:expansion-general}) and (\ref{eq:solution}), the mentioned scalar for $\lambda=1$ becomes
\begin{equation}\label{eq:Salar4}
\left(\dot\Theta+\frac{1}{3}\Theta^2\right)|_{\lambda=1}=\frac{\omega ^2 \hbar ^2 \left(9 (\gamma +2)^2 M^2-12 (\gamma +2) M r+8 r^2\right)}{12 c^2 m^2 r^2 (r-(\gamma +2) M)^2}.
\end{equation}
One can examine its dependence on $\omega$ in the solar system scale, as we have shown in the previous section. It can be observed form Fig.~\ref{fig:Quantum-lambda=1} that for any typical value of $\omega$, the divergence is guaranteed. In this case, the gravitons are in need of attaining very small values of energy. However everything changes once we turn to the strong gravitational effects on gravitons. In this case, namely for $\lambda=4$, the scalar $\left(\dot\Theta+\frac{1}{3}\Theta^2\right)|_{\lambda=4}$ becomes rather complicated and is indeed of order 20 in distance. As in Fig.~\ref{fig:Quantum-lambda=4} which has been plotted for the cosmological range we considered in the previous section, no quantum potential can confine gravitons to exist on diverging congruences; they will eventually converge. It is observed that they obtain much greater values of energy. 

\section{Summary of the Results}\label{sec:conclusion}

One if the essential features of QRE is that it prevents any congruence of massive gravitons from crossing. On the other hand, all theories with massive gravitons agree upon extremely weak couplings of matter and gravity, which becomes even completely negligible inside associated Compton wavelength. In this paper we examined one Lorentz-violating massive theory by finding the quantum  potential, needed by the graviton congruence to avoid convergence. Being rather sensitive to the parameter $\lambda$, the spherically symmetric background spacetime was applied by splitting up to two distinct categories $\lambda=1, 4$ which inspections revealed that divergence becomes possible when the test particles' energies do not exceed those in the asymptote. For the case of $\lambda=1$ and under the Solar system conditions, it became clear that as it was expected, very small amounts of quantum potential can provide permanent divergence. Such values show that the applied massive theory of gravity with $\lambda=1$ is totally consistence with QRE, even in the regions inside the associated Compton wavelength. However for the case of $\lambda=4$, the QRE results in huge values of quantum potential within such conditions. In fact it turns out that $V_Q^4$ is only reasonable on the Compton wavelength, and inside, it is not. Hence, one can infer that the applied massive theory for $\lambda=4$ will indeed break down its integrity with the QRE since it shows strong coupling to matter. One reason, appears to be because of the strong gravitational field in the case of $\lambda=4$, provided by the black hole solution in Eq.~(\ref{eq:solution}). In such situations, we can no longer consider linearized couplings of matter fields and gravitational action. In fact the vigorous gravitational force violates weak coupling between gravity and matter. Because of such conditions, the massive theory fail to conform to the QRE for higher values of $\lambda$. This was also shown through another approach, where the gravitons were considered to be initially bounded by a confined Hamilton-Jacobi condition, to the quantum potential. This provided us a relativistic quantum viewpoint for bosons like graviton which were regarded as a wave function described on a fixed background. This way, we discussed the validity of diverging non-geodesic time-like congruences in both cases of $\lambda=1,4$. There, once again, it appeared that the strong gravitational effect will break the conformity with the necessity of divergence, because the $\lambda=4$ case shows permanent convergence. It should be noted that, even though the Hamilton-Jacobi equation for free particles may confirm possible divergence in large scales, however it is not still the case once the particles are supposed to originate from a quantum potential and move on non-geodesic trajectories. In conclusion, if QRE is supposed to be exploited to investigate congruence evolutions in black hole regions, one should be aware of possible extensions and the capabilities of the gravitational theory, in generating extremely strong fields.

\end{document}